\documentstyle[12pt,aasms4,epsf]{article}
\newcommand{\ha}{H$\alpha$ {}}

\newcommand{\hb}{H$\beta$ {}}

\newcommand{\hii}{H~\small{II}\normalsize~ }

\newcommand{\ana}{{A\&A}}
\newcommand{\anasp}{{A\&AS}}
\begin{document}  
\pagenumbering{arabic}
\title {\bf{UV Colors and Extinctions of HII Regions in the 
Whirlpool Galaxy (M51)}}
\author{ Jesse K. Hill\altaffilmark{1,5}, 
William H. Waller\altaffilmark{1,5},
Robert H. Cornett\altaffilmark{1,5},
Ralph C. Bohlin\altaffilmark{2}}
\author{K.-P. Cheng\altaffilmark{7},
Susan G. Neff\altaffilmark{5}, 
Robert W. O'Connell\altaffilmark{3},
  Morton S. Roberts\altaffilmark{4}} 
\author{Andrew M. Smith\altaffilmark{5}, 
P. M. N. Hintzen\altaffilmark{5,6}, Eric P. Smith\altaffilmark{5}}
\author{Theodore P. Stecher\altaffilmark{5}}
\altaffiltext{1}{Hughes STX, 4400 Forbes Blvd, Lanham, MD 20706}
\altaffiltext{2}{Space Telescope Science Institute, Homewood Campus, 
   Baltimore, MD 21218}
\altaffiltext{3}{University of Virginia, P. O. Box 3818, 
  Charlottesville, VA 22903}
\altaffiltext{4}{National Radio Astronomy Observatory, Edgemont Rd., 
   Charlottesville, VA 22903}
\altaffiltext{5}{Laboratory for Astronomy and Solar physics, 
NASA/GSFC, Greenbelt, MD 20771}
\altaffiltext{6}{California State University, Dept. of Phys. \& Astron.,
Long Beach CA 90840}
\altaffiltext{7}{California State University, Dept. of Phys.,
Fullerton CA 92634}
\pagestyle{plain}
\vspace{.5in}
\begin{abstract}
Far-UV (wavelength $1520$ \AA), U, \ha, and R 
images of the interacting Sbc spiral galaxy M51 were obtained 
by the Ultraviolet Imaging Telescope (UIT) during the {\sl Astro-2}
Spacelab mission of 1995 March and at Mt. Laguna Observatory.
The $\mu_{152}-\mu_{U}$ radial 
gradient of over a magnitude, becoming bluer with increasing
radius, is attributed primarily 
to a corresponding radial extinction gradient.
Magnitudes in both UV bands and \ha fluxes are reported for $28$
\hii regions.
Optical extinctions for the $28$ corresponding UV sources
are computed from the measured $m_{152}-U$ colors
by fitting to the optical extinctions of Nakai \& Kuno
\markcite{nakaia} (1995).
The normalized far-UV extinction $A_{152}/E(B-V)$
increases with increasing galactocentric distance
or decreasing metallicity, from $5.99$ to $6.54$, compared
with the Galactic value $8.33$.
The best-fit $m_{152}-U$ color for no extinction, $-3.07$,  
is the color of a model solar metallicity
starburst of age $\sim 2.5$ Myr with IMF slope $-1.0$. 
HII regions show decreasing observed \ha fluxes
with decreasing radius, relative to the \ha fluxes 
predicted from the observed $f_{152}$
for age $2.5$ Myr, after the \ha and
$f_{152}$ are corrected for extinction.
We attribute the increasing fraction of ``missing'' \ha flux with
decreasing radius to increasing extinction in the Lyman continuum.
The increasing extinction-corrected far-UV flux of the \hii regions 
with decreasing distance to the nucleus is probably a result of the 
corresponding increasing column density of the interstellar gas
resulting in
larger mass OB associations.
The estimated dust-absorbed Lyman continuum flux
is $\sim 0.6$ times the 
far-infrared energy flux of M51 observed by IRAS. 
 
\end{abstract}

\clearpage 
\section{Introduction}
Because of its proximity, almost face-on orientation, and 
prominent spiral structure,
the interacting Sbc galaxy M51 has been a favorite target of observational
astronomers for many years.
The Ultraviolet Imaging Telescope image of M51 reported here,
obtained with $\sim 3$ arcsec resolution,
represents a 5-fold increase in resolution compared to prior UV
imaging (Bohlin et al. \markcite{bohlina} 1990a; Bersier
\markcite{bersiera} et al. 1994). 
The image was obtained during the
Astro-2 spacelab mission of 1995 March.
Ultraviolet images of spiral galaxies are dominated by the OB/\hii
complexes in the spiral arms (Hill, Bohlin \& Stecher \markcite{hilla} 1984;
Bohlin et al. \markcite{bohlina} 1990a, \markcite{bohlinc} 1990b; 
Hill et al. \markcite{hillb} 1992).
Complementary CCD images in the U, \ha, and R bands
were obtained at Mt. Laguna Observatory to better investigate the
star-forming complexes. 
A log of the observations used in this investigation is given in 
Table~\ref{obslogtable}.

UIT is a 38 cm Ritchey-Chr\'{e}tien telescope, which images a $40'$ diameter
field to the cathode of a two-stage magnetically 
focussed image intensifier, coupled by fiber optics to 
II-aO film.
The intensifier has a UV-sensitive, solar-blind CsI
photocathode.
(The intensifier for the near-UV camera, which functioned well during
the 1990 Astro-1 mission, was apparently damaged during the Astro-2 launch,
so near-UV data were obtained from the ground in the U band.)
The UIT instrument, the available bandpasses, and the 
reduction of the digitized film
images to arrays of flux-calibrated integer pixels are discussed 
by Stecher et al. \markcite{stechera} (1992).
The bandpass used in this investigation is termed
B1, with centroid $1520$ \AA\ and width $354$ \AA.
Fluxes $f$ measured in this UIT band are converted to magnitudes
$m_{152}$ by the relation $m_{152}=-2.5\times \log f-21.10$.
The longest B1 exposure obtained, and the one used in this
investigation, is of duration $1100.5$ s.  
A north-up, east-left version of the image was corrected for 
image distortion using the method of Greason et al. \markcite{greasona} (1994) 
and used for subsequent analysis. 

A UIT precursor aboard a sounding-rocket imaged M51 at wavelength
2250 \AA, with $15$ arcsec resolution (Bohlin et al. \markcite{bohlina} 1990a).
The most prominent features in M51 at 2250 \AA\ are 
a region surrounding the nucleus, of size $\sim 1$ arcmin and the
group of bright \hii regions about $2$ arcmin NE from the nucleus.
The interacting companion NGC 5195 is much less prominent in the near-UV than at
optical wavelenths.  The nuclear region shows a complex structure,
with evidence for star formation and possibly for an Inner Lindblad Resonance
at radius $\sim 10-15$ arcsec.
In combination with a digitized photographic image in the $U$ band, evidence
was found for a variation in $m_{225}-U$ colors of bright \hii regions of about
$1$ magnitude (here, $m_{225}$ is the magnitude in the sounding rocket band).

A UV image of M51 at wavelength 2000 \AA, also with $15$ arcsec resolution,
was obtained by Bersier et al. \markcite{bersiera} (1994) with a balloon-borne 
telescope. 
Bersier et al. found that the UV emission in the spiral arms usually
peaks about $7$--$11$ arcsec in the direction of galactic 
rotation from the \ha emission. 
They estimated star formation rates and star formation efficiencies. 

\begin{figure}[htbp]
\figurenum{1a}
\epsffile{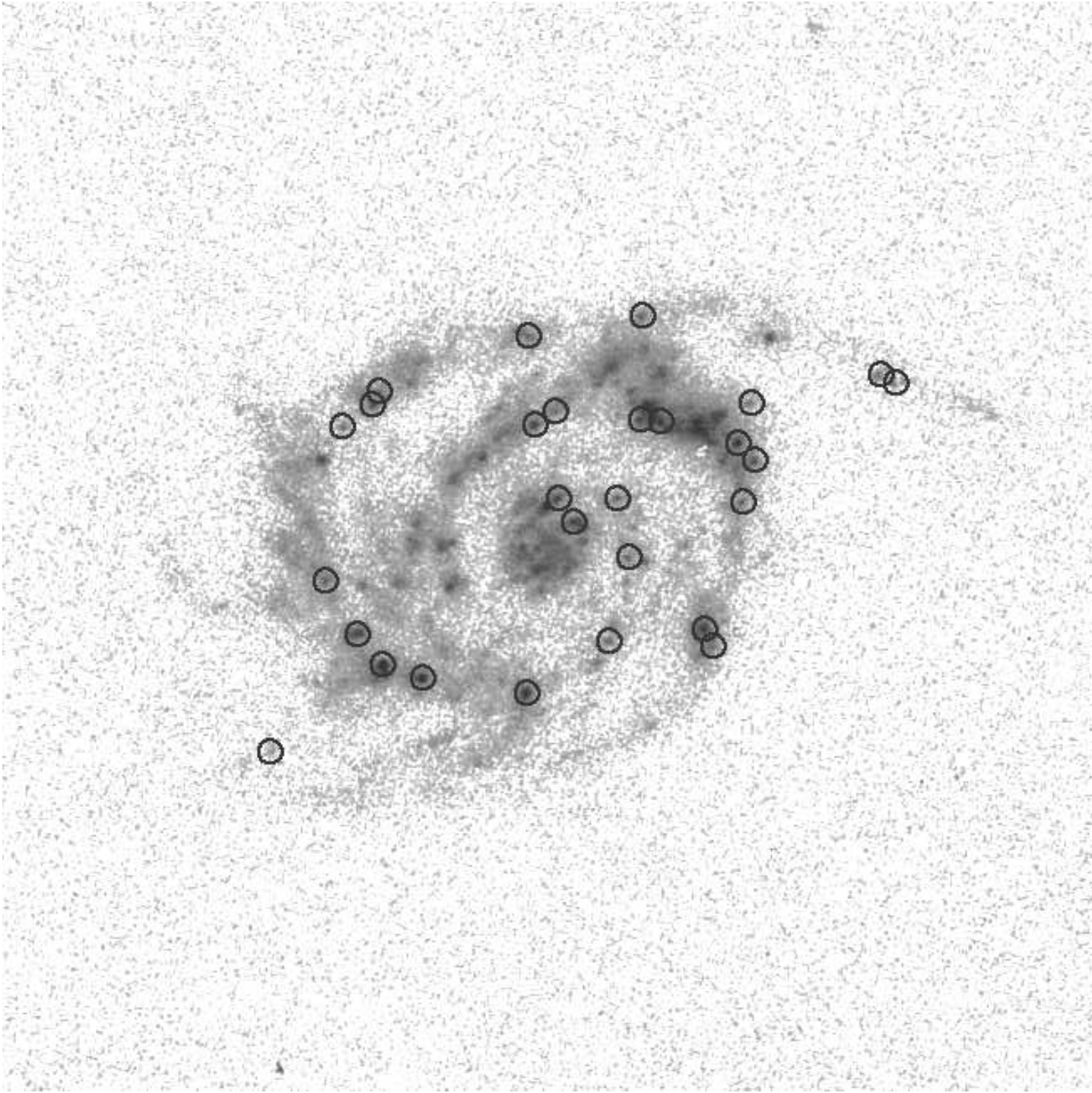}
\caption{Far-UV UIT B1 band image FUV2545 of M51.  North is up, 
east is to the left.  The 28 \hii regions discussed are circled.
\label{fuvimgfig}}
\end{figure}

\begin{figure}[htbp]
\figurenum{1b}
\epsffile{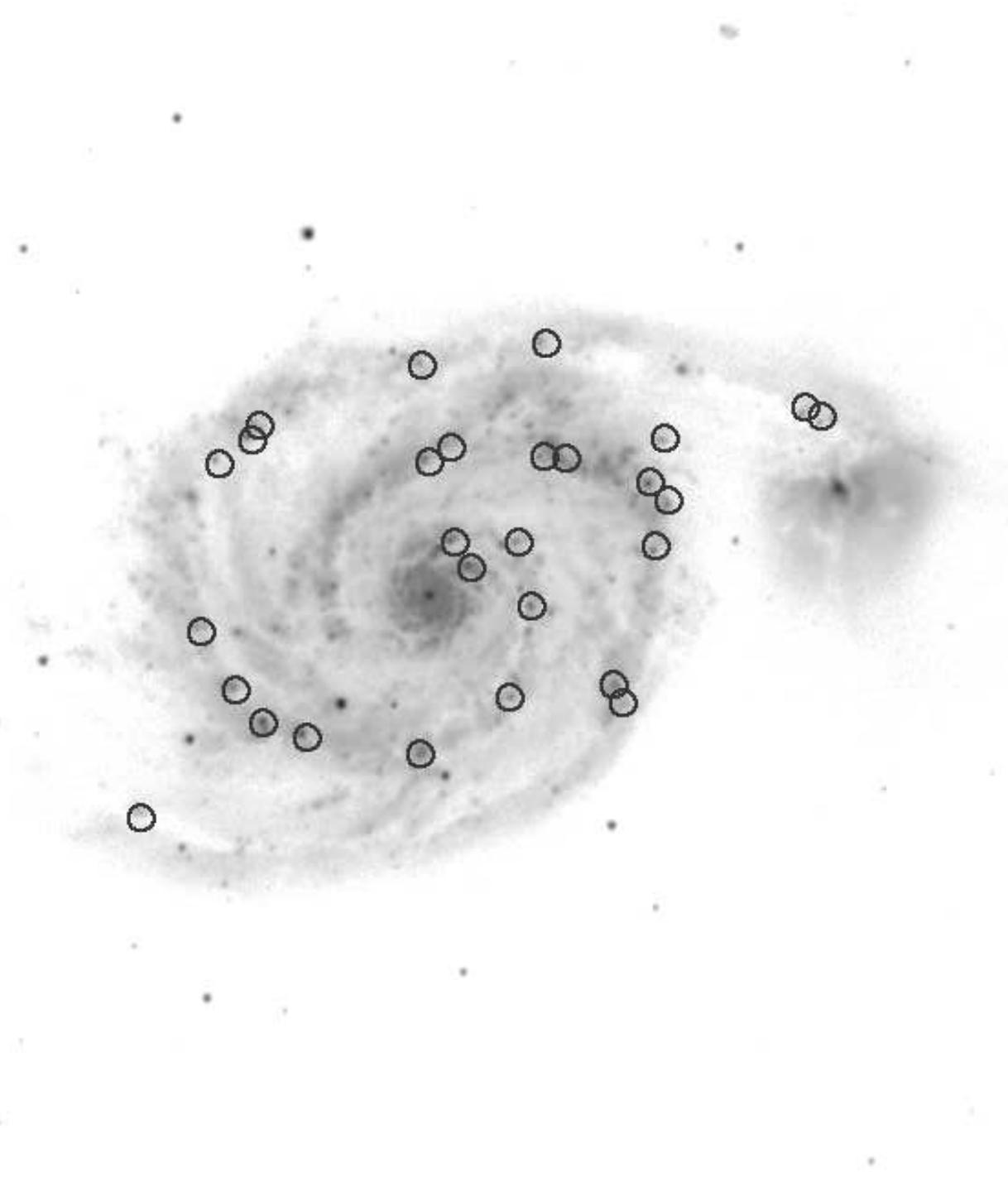}
\caption{Near-UV UIT U band image of M51, registered with 
Fig.~\ref{fuvimgfig}.
The 28 HII regions discussed are circled.\label{nuvimgfig}}
\end{figure}

\begin{figure}[htbp]
\figurenum{1c}
\epsffile{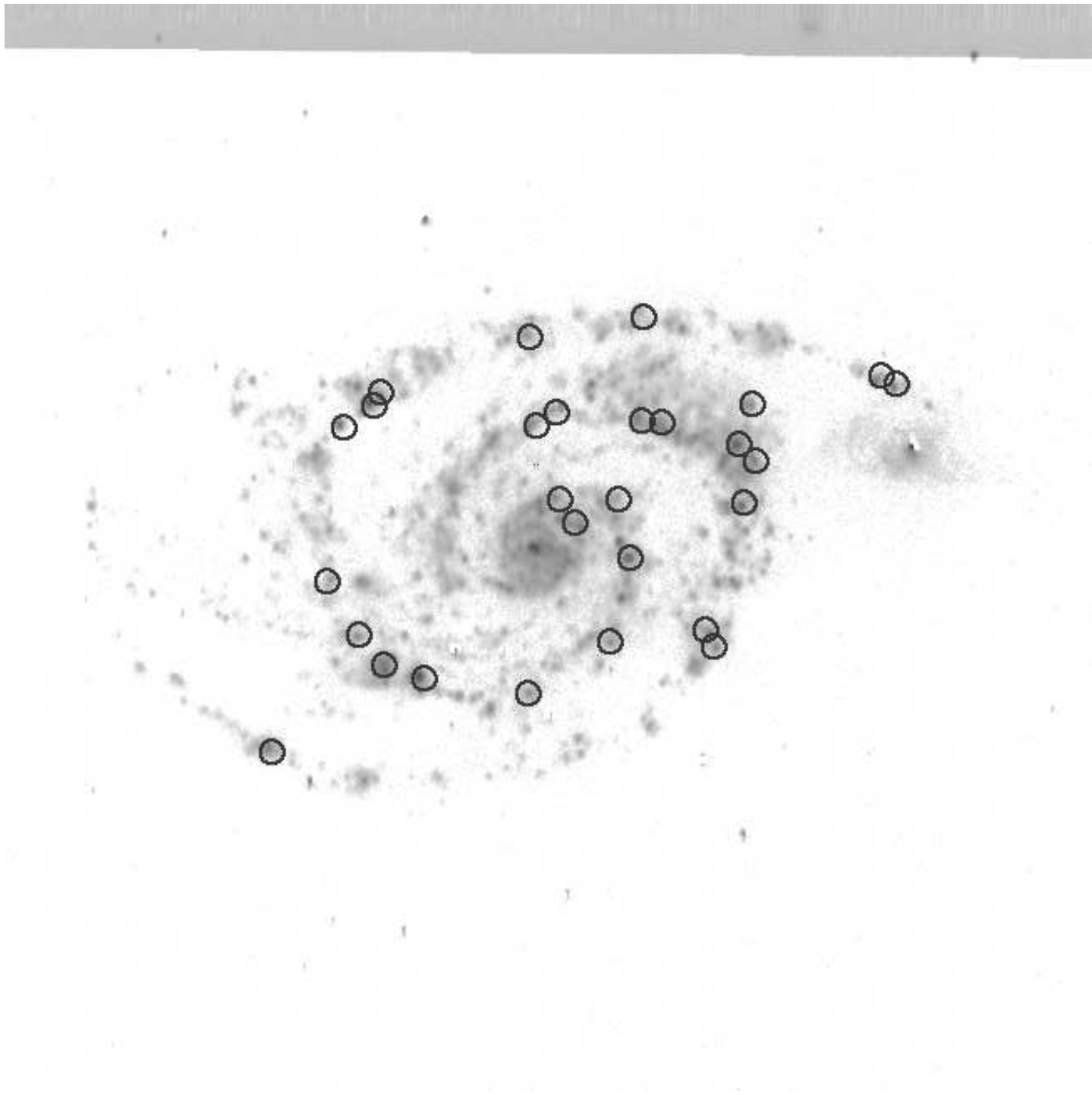}
\caption{Continuum-subtracted \ha image of M51, registered with 
Fig.~\ref{fuvimgfig}.  The 28 HII regions discussed are circled.
\label{haimgfig}}
\end{figure}

CCD images of M51 were obtained in the Johnson U band, the Kron-Cousins R band,
and at \ha 
with the Mt. Laguna 1.0 m using an interference filter of bandwidth 61\AA.
Figures~\ref{fuvimgfig}--\ref{haimgfig} (Plates XX-XX) show the UIT far-UV 
image, 
the U image, and the \ha image, after subtraction of the red
continuum according to the formulations in Waller (1990).
The companion galaxy NGC 5195, visible on the U band
image, is not detectable on the UIT far-UV image.
The sounding rocket image appears as Figure~1b in Bohlin et al.
\markcite{bohlina} (1990a).
We present and discuss multiband photometry of 28
M51 UV sources,
and surface photometry in the far-UV and U band images.
The 28 UV sources investigated 
are circled on Figures~\ref{fuvimgfig}--\ref{haimgfig}.  

\section{The UV Color profile and Source Photometry}
From a Fourier analysis of their UV image, 
Bersier et al.\markcite{bersiera} (1994) determined the position angle
and inclination of M51 to be 26 degrees and 42 degrees, respectively, 
consistent with previous
estimates.
We adopt their position angle and inclination to contruct
equivalent face-on azimuthally averaged radial profiles in the UIT
B1 band and the ground-based U band.  The UV profiles are
similar to the profile
of Bersier et al. \markcite{bersiera} (1994) and the profile of 
Bohlin et al. \markcite{bohlina} (1990a).
From the B1 and U band profiles the $\mu_{152}-\mu_{U}$ 
color profile is constructed
and plotted in Figure~\ref{fuvufig}.  
A UV color gradient of about $2$ magnitudes is found from the nucleus
to the outermost measured UV sources, at radius 11 kpc.

We identify 28 spiral arm sources on the UIT image
with average separation $2.6$ arcsec from the corresponding \hii region
on the \ha image, coincident
within the estimated accuracy of the distortion correction applied to
the UIT images (Greason et al. \markcite{greasona} 1994).
Aperture photometry is performed of these sources on the UV and \ha images
with apertures of radius
$8$ arcsec ($370$ pc for distance $9.6$ Mpc).
The sky value is determined as the mode of pixels within annuli
with inner and outer radii $11$ and $22$ arcsec. 
Aperture photometry
is performed using IDL implementations of 
DAOPHOT software (Stetson \markcite{stetsona} 1987).
No aperture correction is employed in the UIT image photometry because
any extended emission beyond the boundary of the $8.0$ arcsec aperture
appears to be associated with extended, diffuse spiral arm emission,
rather than the source itself.  The \ha aperture correction for the 
nearly-isolated \hii region Rand 454, is only about $ 25$\%.  Hence,
we apply no aperture corrections to the \ha fluxes either, because the
corrections are small, and also because the sources do not have identical
profiles.  

{\sl IUE} spectra of isolated stars from several 
fields observed by UIT are utilized in
determining the UIT calibration,
thought to be accurate to about 15\% relative to {\sl {IUE}}.
UIT instrumental fluxes used in determining the calibration 
are from apertures large
enough to contain all the stellar flux.
Calibration of the U band photometry, with uncertainty about 5\%,
is based on the photoelectric photometry
of Schweizer \markcite{schweizera} (1976).

Subtraction of the red continuum from the \ha band image
is accomplished
by scaling count rates of stars in the R image to equal their
counterparts in the \ha image (Waller \markcite{wallera} 1990).
The effect of line emission in the R-band image is estimated
and removed using the filter curves for the two bands. 
The contribution of [NII] ($\lambda\lambda 6548, 6584$) line emission
to the total \ha+[NII] line flux is estimated to be 35\% in 
emission (McCall \markcite{mccalla} et al 1985) and 28 \% through the 
\ha filter.
McCall et al.\markcite{mccalla} (1985) have determined that the ratio 
[NII]/\ha
is independent of excitation over a range $-0.35$ to $-0.56$ in
$\log\ ([OII]+[OIII])$/\hb.
The quoted \ha fluxes have been corrected accordingly.
Absolute calibration of the \ha and R images uses CCD images 
of the spectrophotometric standard star BD +08 2015
taken on the same photometric night with the same instrument.
The resulting \ha photometry of the \hii region Rand 454
agrees with that of Rand \markcite{randa} (1992) to within 8\%.

UV magnitudes and \ha fluxes for the $28$ sources are given
in Table~\ref{hiiregtable}.
Table~\ref{hiiregtable} contains the source index number (col. 1),
the \hii region number from Rand \markcite{randa}(1992) (col. 2),
the distance in arcsec east of the nucleus (col. 3), 
the distance in arcsec north of the nucleus (col. 4), 
the far-UV magnitude ${m_{152}}$ (col. 5), the estimated error in the far-UV
magnitude (col. 6), the U magnitude (col. 7), the estimated
error in the U magnitude (col. 8), the \ha flux (col. 9), the
estimated fractional error in the \ha flux (col. 10), and the $A_{V}$
computed from the UV color (col. 11).
Figure~\ref{fuvufig}, which plots the large-scale radial variation in the
$\mu_{152} - \mu_{U}$ color, also plots as solid circles the  
$m_{152}-U$ colors of the 28 UV sources.
The colors of the sources are 
$\sim 0.5-1.0$ mag bluer than
the large-scale UV color at the same radius, 
in nearly all cases.
The UV CMD for the \hii regions of M51 is similar to the UV CMD
for the M81 HII regions (Hill et al. \markcite{hillc}1995), except 
that for M81 the UIT
near-UV magnitudes $m_{249}$ were used, rather than ground-based
U.
The brightest of the $28$ sources in the far-UV band has luminosity 
about $800$ times the far-UV luminosity of the Orion Nebula 
(Bohlin et al. \markcite{bohlinb} 1982), while the faintest has
a far-UV luminosity $\sim 80$ times that of Orion. 

\section{Extinctions and Ages of HII Regions}

Evolving star cluster spectra are modeled using the evolutionary 
stellar models of Schaller et al. \markcite{schallera}(1992),
together with the model atmosperes of Kurucz 
\markcite{kurucza} (1992)
for IMF slopes $-1.0, -1.5$, and $-2.0$,
assuming solar metallicity.  For each of these evolving models, $m_{152}-U$
colors are determined as a function of age.
For an IMF slope equal to $-1.0$, 
close to the slope $-1.08$ found by 
Hill et al. \markcite{hilld} (1994) for the OB associations near 30 Dor,
the UV color begins at $\sim -3.2$
at age zero, and evolves to $\sim -2.6\pm{0.16}$ for ages $3-15$ Myr.
Steady-state star formation models predict $m_{152}-U$ colors
$-2.47$, $-2.21$, and $-1.92$ for the three IMF slopes.
Steady-state models with twice solar metal abundances give UV colors
redder by $\sim 0.10$ mag for the three IMF slopes.
The relatively modest change in predicted UV color for a factor two
increase in metallicity suggests that the primary cause of the
observed color gradient in M51 is a radial extinction gradient.

Nakai \& Kuno \markcite{nakaia} (1995) revised the $A_{V}$ for the \hii
regions observed by van der Hulst et al. \markcite{hulsta} (1988), 
taking into account
the variation in \hii region gas temperature caused by the 
effect of the metallicity
gradient on the adundance of coolants.
A linear relation is fit between the revised $A_{V}$ of Nakai \&
Kuno and the $m_{152}-U$ colors measured here, for the $14$ \hii
regions in common between the two lists. 
The fit relation between the $A_{V}$ of Nakai \& Kuno 
\markcite{nakaia} (1995) and the UV colors is then
used to estimate $A_{V}$ for all 28 \hii regions.
We estimate the metallicities of the \hii regions from the
galactocentric distance, assuming that the \hii region Rand 454
has solar metallicity and that the 
metallicity variation with increasing 
distance is $-0.14$ dex/arcmin (Villa-Costas \& Edmunds
\markcite{vilacostasa} 1992).

The UV spectrum over the wavelength interval $400$ to $4000$ \AA\ is computed
for IMF slope $-1.0$ (Hill et al. \markcite{hilld} 1994), age $2.5$ Myr, 
and metallicities
equal to the solar metallicity and twice the solar metallicity, using the
evolutionary models of Schaller et al. \markcite{schallera}(1992).
For twice solar metallicity, 
the predicted ratio of Lyman continuum flux to the flux at 1500 \AA\
is only half that predicted for solar metallicity.  
The spectra of the innermost \hii regions, whose
metallicities are estimated at $\sim 4$ times solar, 
are extrapolated from the $2$ times solar metalicity model.
For them, the predicted ratio of Lyman continuum flux to the flux at 1500 \AA\
is only $0.3$ times that predicted for solar metallicity.  
UV colors $m_{152}-U$ are computed from these model
spectra, and used to determine the variation of the unreddened UV color
with radius, allowing computation of the UV color excess
$E(m_{152}-U)$ for each of the \hii regions.
The ratio of the UV color excess to $E(B-V)$ is then used 
to derive far-UV extinctions, by interpolating the normalized far-UV 
extinction over the normalized UV color excess between
the 30 Dor nebular extinction curve (Fitzpatrick \& Savage
\markcite{fitzpatrickb} 1984),
and the Orion extinction curve (Bohlin \& Savage 
\markcite{bohlind} 1981).
We derive normalized far-UV extinctions $A_{152}/E(B-V)$ for the
\hii regions which vary from $5.99$ to $6.54$, with
average value $6.26$. 
These normalized far-UV extinctions are plotted versus metallicity 
relative to solar in Figure~\ref{fuvextncfig}. 
The best-fit $m_{152}-U$ color for zero extinction is $-3.07$, 
the color
of an evolving solar metallicity starburst of age $\sim 2.5$ Myr,
approximately the average age obtained by Hill et al. 
\markcite{hillc} (1995) 
for the M81 \hii regions.

The optical extinctions $A_{V}$ estimated from the UV colors 
for the $28$ HII regions
are given in Table~\ref{hiiregtable}, col. 11.
Ages for these sources are estimated to be $\sim 2.5$ Myr.
Because these $28$  \ha sources are
closely coincident with the corresponding far-UV sources, we
can estimate the fraction of the Lyman continuum photons which
are absorbed by HI atoms rather than by
dust grains, as the ratio of the observed \ha flux (corrected
for the extinction estimated from $m_{152}-U$) to the \ha flux
expected from the extinction-corrected $m_{152}$, the evolving starburst
model with IMF slope equal to $-1.0$, and the assumption that the age
is $2.5$ Myr, and taking into account the variation of metallicity
with radius.

We define an effective Lyman continuum extinction as 
$A_{Ly,eff}=2.5\times log(f_{H\alpha,UV}/f_{H\alpha})$,
where $f_{H\alpha,UV}$ is the \ha flux predicted from the
extinction corrected UV flux and the assumed age $2.5$ Myr, and $f_{H\alpha}$
is the extinction corrected \ha flux.
Figure~\ref{lyextncfig} plots the effective Lyman continuum extinction vs. E(B-V)
computed from the estimated V-band extinction.
The plot shows that the two measures of extinction are strongly correlated.
The ratio $A_{Ly,eff}/E(B-V)$ is approximately equal to the 
normalized extinction in M51 at a typical Lyman continuum wavelength,
times the fraction of the E(B-V) which takes place inside the \hii
region.
If the normalized extinction at Lyman continuum wavelengths
in M51 is approximately equal to $19.0$, the value obtained by
Clayton et al. (1996) for two lines of sight in the LMC, then
for the 28 \hii regions investigated here, the fraction 
of the optical extinction which takes place inside the \hii region averages
$0.15$, and ranges from $0.07$ to $0.23$.

We define ``missing'' \ha flux as
the difference between the \ha flux predicted by the corrected UIT flux
and the corrected \ha flux.
The estimated ratio of the 
``missing'' Lyman continuum energy flux (equal in photon flux to $\sim$ twice
the ``missing'' \ha flux), 
to the total observed IRAS energy flux is $0.60$.

The extinction-corrected far-UV magnitudes of the 
UV sources become brighter with
decreasing distance from the nucleus.  The masses
of the OB associations exciting the \hii regions are expected be proportional
to the corrected far-UV fluxes.  The interstellar gas column densities 
exhibit a similar trend (Nakai \& Kuno \markcite{nakaia} 1995).  The CO 
azimuthal profiles of
Kuno et al. \markcite{kunoa} (1995) show larger peak fluxes above background
in the integrated intensity for
decreasing radii, which is consistent with an increase in the masses 
of the OB associations
exciting the \hii regions with decreasing
galactocentric radius.
  
\section{Acknowledgements}
We gratefully acknowledge the contributions made by the
many NASA personnel involved in the {\em Astro-2} mission.
We also thank IPAC personnel for providing the 
HIRES IRAS images
of M51 used in this investigation.
 
Funding for the UIT project has been through the Spacelab Office
at NASA headquarters under Project number 440-51.  
RWO gratefully acknowledges NASA support of portions of this research
through grants NAG5-700 and NAGW-2596 to the University of Virginia.  
%
%
\clearpage
\begin{center}
\begin{deluxetable}{llllll}
\tablecaption{Log of Observations of M51\label{obslogtable}}
\tablehead{\colhead{Observatory} & \colhead{Telescope} & \colhead{band} & 
\colhead{wavelength (\AA)} & \colhead{bandwidth (\AA)} & 
\colhead{exposure time (s)}}
\startdata
{\sl Astro-2} & UIT     & B1   & 1520 & 354 & 1100.5 \nl
Mt. Laguna  & 1m      & U  & 3650 & 700 & 1200.0 \nl
Mt. Laguna  & 1m      & \ha  & 6573 & 61 & 600.0 \nl
Mt. Laguna  & 1m      & R  & 6500 & 1500.0 & 1500.0 \nl
\enddata
\end{deluxetable}
\end{center}
\clearpage
%
%
\begin{deluxetable}{rrrrrrrrrrr}
\tablecaption{HII Regions\label{hiiregtable}}
\tablehead{\colhead{Ind} & \colhead{Rand} & \colhead{X} & \colhead{Y} & 
\colhead{$m_{152}$} & \colhead{err} & \colhead{U} & \colhead{err} & 
\colhead{$f_{H\alpha}\times 10^{14}$} & \colhead{frac. err} & \colhead{$A_V$}}
\startdata
1 &190 & 158.1 & 68.5 & 15.08 & 0.08 & 16.46 & 0.08 & 3.07 & 0.93 & 2.41  \nl
2 &173 & 144.6 & -7.7 & 14.76 & 0.04 & 16.38 & 0.06 & 7.61 & 0.37 & 2.06 \nl
3 &218 & 118.2 & 230.0 & 15.30 & 0.06 & 17.33 & 0.12 & 8.78 & 0.32 & 1.48 \nl
4 &219 & 112.5 & 240.0 & 15.47 & 0.07 & 17.97 & 0.22 & 7.35 & 0.38 & 0.81 \nl
5 &147 & 106.6 & -108.8 & 14.64 & 0.05 & 16.05 & 0.05 & 17.69 & 0.16 & 2.36 \nl
6 &321 & 99.1 & 142.5 & 14.85 & 0.08 & 16.45 & 0.06 & 8.37 & 0.34 & 2.10 \nl
7 &146 & 97.8 & -113.2 & 14.63 & 0.04 & 15.97 & 0.05 & 25.60 & 0.11 & 2.46 \nl
8 &275 & 93.6 & 10.3 & 15.25 & 0.08 & 16.09 & 0.06 &  6.00 & 0.47 & 3.18 \nl
9 &307 & 87.8 & 67.8 & 14.03 & 0.04 & 15.45 & 0.04 & 14.68 & 0.19 & 2.35 \nl
10&309 & 86.2 & 82.0 & 13.75 & 0.03 & 15.22 & 0.03 & 19.73 & 0.14 & 2.29 \nl
11 &270 & 84.2 & -4.1 & 14.59 & 0.05 & 15.62 & 0.04 & 9.78 & 0.29 & 2.90 \nl
12 &142 & 82.5 & -133.6 & 16.24 & 0.20 & 17.49 & 0.15 & 7.49 & 0.38 & 2.60 \nl
13 &320 & 72.0 & 134.3 & 13.94 & 0.05 & 15.40 & 0.06 & 28.05 & 0.10 & 2.30 \nl
14 &323 & 59.6 & 145.4 & 14.59 & 0.07 & 15.66 & 0.05 & 13.59 & 0.21 & 2.85 \nl
15 &24 & 34.5 &  52.4 & 15.35 & 0.08 & 15.99 & 0.05 & 5.23  & 0.54 & 3.46 \nl
16 &9 & 34.3 &  12.0 & 14.10 & 0.04 & 15.08 & 0.03 & 5.20 & 0.55 & 2.98 \nl
17 &332 & 31.8 & 136.9 & 14.72 & 0.06 & 16.08 & 0.06 & 15.93 & 0.17 & 2.44 \nl
18 &13 & 18.3 & 22.6 & 14.23 & 0.06 & 14.94 & 0.03 & 5.43 & 0.54 & 3.37 \nl
19 &32 & -6.0 & 59.6 & 15.32 & 0.07 & 15.96 & 0.05 & 21.20 & 0.13 & 3.47 \nl
20 &96 & -22.0 & -145.7 & 15.07 & 0.06 & 16.36 & 0.06 & 4.28 & 0.66 & 2.54 \nl
\tablebreak
21 &401 & -54.5 & 111.1 & 13.93 & 0.02 & 15.34 & 0.03 & 11.76 & 0.24 & 2.36 \nl
22 &80 & -57.7 & -123.7 & 14.41 & 0.04 & 15.77 & 0.04 & 6.41 & 0.44 & 2.43 \nl
23 &45 & -62.7 & 46.6 & 15.25 & 0.06 & 16.22 & 0.06 & 7.12 & 0.40 & 2.99\nl
24 &403 & -66.2 & 116.7 & 14.58 &0.05 & 15.93 & 0.05 & 7.96 & 0.35 & 2.45 \nl
25 &79 & -78.8 & -106.5 & 13.74 & 0.02 & 14.95 & 0.03 & 30.29 & 0.09 & 2.65\nl
26 &... & -87.9 & -79.9 & 14.15 & 0.03 & 15.28 & 0.04 & 36.52 & 0.07 & 2.77 \nl     
27 &61 & -98.0 & -9.4 & 14.17 & 0.03 & 15.45 & 0.04 & 6.34 & 0.45 & 2.55 \nl      
28 &455 & -137.8 & -182.4 & 16.09 & 0.14 & 17.36 & 0.11 & 14.76 & 0.19 & 2.58 \nl  
\enddata
\end{deluxetable} 
%
%

%
%
\clearpage

\stepcounter{figure}

\begin{figure}[htbp]
\epsffile{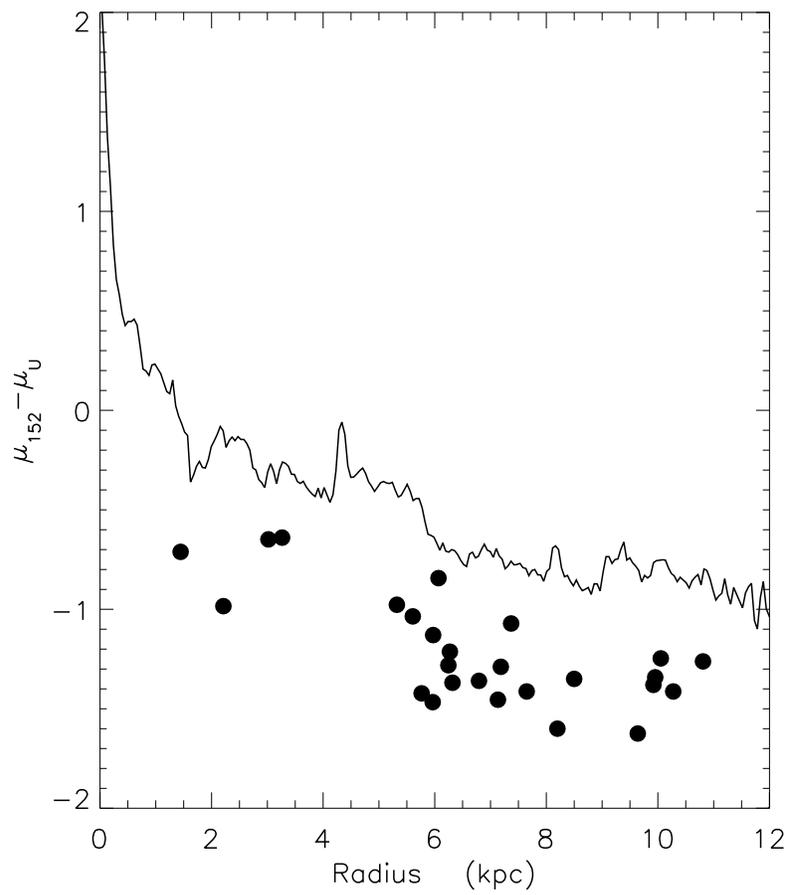}
\caption{M51 $\mu_{152}-\mu_U$ radial profile.  The 
28 UV sources are plotted as solid circles.\label{fuvufig}}
\end{figure}

\begin{figure}[htbp]
\epsffile{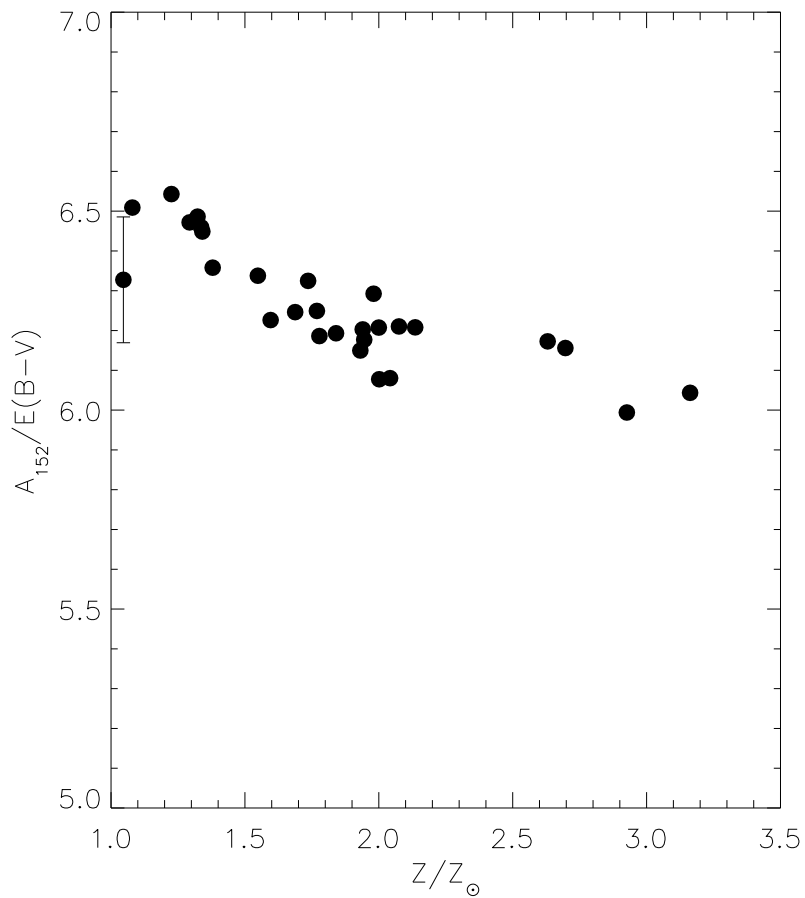}
\caption{The normalized far-UV extinction $A_{152}/E(B-V)$ plotted 
against the metallicity Z relative to the solar metallicity.
The largest metallicities occur at the smallest galactocentric
radii.\label{fuvextncfig}}
\end{figure}

\begin{figure}[htbp]
\epsffile{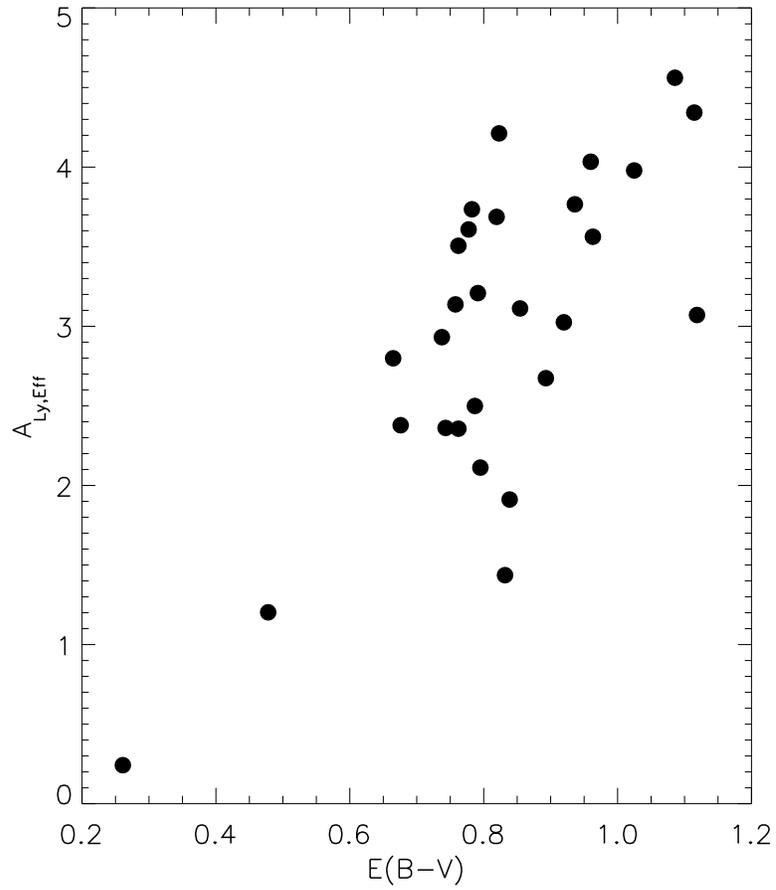}
\caption{The effective Lyman continuum extinction, computed
as described in the text, is plotted vs. E(B-V) fit
from $m_{152}-U$.\label{lyextncfig}}
\end{figure}
%
%
\end{document}